\documentclass{aa}  
\usepackage{graphicx}
\usepackage{txfonts}

\usepackage{lscape}
\usepackage{aas_macros} 
\usepackage{amssymb}
\usepackage{natbib}  
\bibpunct{(}{)}{;}{a}{,}{,} 
\usepackage{longtable}
\usepackage{graphicx}
\usepackage[hang,bf,footnotesize]{subfigure}
\usepackage[figuresright]{rotating}

\begin{document}
\titlerunning{Spatial distribution of the circumstellar disks in NGC~6611}	
   \title{Correlation between the spatial distribution of circumstellar disks and massive stars in the open cluster NGC~6611\thanks{Based on observations made with the European Observatory telescopes obtained from the ESO/ST-ECF Science Archive Facility.}.}

   \subtitle{Compiled catalog and cluster parameters.}

   \author{M. G. Guarcello\inst{1,2} \and L. Prisinzano\inst{2} \and G. Micela\inst{2} \and F. Damiani\inst{2} \and G. Peres\inst{1} \and S. Sciortino\inst{2}}

   \offprints{mguarce@astropa.unipa.it}

   \institute{Dipartimento di Scienze Fisiche ed Astronomiche, Universit\'a di Palermo, Piazza del Parlamento 1, I-90134 Palermo Italy
   \and
   INAF - Osservatorio Astronomico di Palermo, Piazza del Parlamento 1, 90134 Palermo Italy}

  \date{}
 
  \abstract
   {The observation of young stars with circumstellar disks suggests that the disks are dissipated, starting from the inner region, by the radiation of the central star and eventually by the formation of rocky planetesimals, over a time scale of several million years. It was also shown that strong UV radiation emitted by nearby massive stars can heat a circumstellar disk up to some thousand degrees, inducing the photoevaporation of the gas. This process strongly reduces the dissipation time scale.}
   {We study whether there exists a correlation between the spatial distribution of stars with circumstellar disks and the position of massive stars with spectral class earlier than B5, in the open cluster NGC~6611.}
   {We created a multiband catalog of the cluster, down to V$\sim$23$^m$, using optical data from a WFI observation at 2.2m of ESO in the BVI bands, the 2MASS public point source catalog and an archival X-ray observation made with CHANDRA/ACIS. We selected the stars with infrared excess (due to the emission of a circumstellar disk) using suitable color indices independent of extinction, and studied their spatial distribution.}
   {The spatial distribution of the stars with K band excess (due to the presence of a circumstellar disk) is anti correlated with that of the massive stars: the disks are more frequent at large distances from these stars. We argue that this is in agreement with the hypothesis that the circumstellar disks are heated by the UV radiation from the massive stars and photoevaporated.}
   {}


   \maketitle
%

\section{Introduction}
	
	Circumstellar disks are important features of star formation process for many reasons. For example, they regulate the angular momentum evolution of the stars and can evolve in planetary systems. Disks form during the gravitational contraction of the protostellar nebula and are dissipated over time scale of several millions years (see, for example, \citealt{Hai01}) . \par
	 When it is not possible to directly observe the circumstellar disks, as in the nearest star-forming regions, the disk dust emission in the infrared bands is used to detect and study them. However, observations show a decline in the emission of the inner region of the disks (mostly NIR and mid-infrared bands) between $\sim$1 and $\sim$10 Myr (see, for example, \citealt{Sici05,Cla05}). This decline can be explained as the normal evolution of the disks, disrupted first in the inner region by the radiation of the central star and by the formation of rocky planetesimals, that are thought to occur fastest in the inner disks \citep{Podo93}. Stars with transition disks that are not characterized by NIR emission but that have strong mid- and far-infrared emission have been observed. \par
	The evolution of circumstellar disks, however, depends strongly on the environment where they formed. It was observed that strong UV radiation (mostly the far UV, with photons with energy between 6.0 eV and 13.6 eV) from the massive stars can heat the temperature of circumstellar disks up to thousands of degrees, inducing the photoevaporation of the gas \citep{Holle94,John98,Sto99,Adams04}. The disk mass loss rate is independent of the distance from the central stars \citep{Holle94} and since the surface density decreases with increasing distance from the central star \citep{Prin81}, the disks are disrupted starting from the outer regions. So the evolution and the thermal structure of disks near massive stars is very different from those of isolated disks. For example,  \citet{John98} shown that the outer region of a circumstellar disk with a mass equal to 0.2 solar masses and a distance less than $10^{17}$ cm from the star $\Theta^1$ Ori (spectral class O6) is disrupted in 10$^6$ years up to a radius of about 1 AU from the central star. \par
	To investigate such effects, we study the stars with circumstellar disks in the young open cluster NGC~6611. This cluster is characterized by a population of pre-main sequence stars younger than 3 million of years (see Sect. \ref{parameter}) and by a large number of massive stars (56 stars with spectral class earlier than B5, \citealt{Hille93}). The massive stars are distributed irregularly in the region, therefore the UV flux is non-uniform in the cluster. NGC~6611 is an ideal target to study the effects of UV radiation on the evolution of the circumstellar disks of low mass stars. The cluster is also within the Eagle Nebula that strongly absorbs the radiation of background stars. This will be very helpful to estimate the distance of the cluster and to avoid all the problems related to contaminating background stars. \par
	\citet{Hille93} (thereafter H93) estimated the age of the pre-main sequence stars of NGC~6611 as between 0.25 and 1 million of years, the cluster distance as 2.0 $\pm$ 0.1 kpc and derived an anomalous extinction law, with R=3.75. \citet{Beli99} (thereafter B99) derived a distance of 2.14 $\pm$ 0.10 kpc, while smaller distance values are reported in the more recent Catalog of Galactic Open Clusters of \citet{Khar05} (1.7 kpc) and in the 2MASS study of \citet{Bona06} (1.8 $\pm$ 0.5 kpc). \par
	We generated a multiband catalog using optical, near-infrared and X-ray data. Optical data will be used to determine the photospheric properties of the stars; NIR data will give indications of the presence of circumstellar disks; X-ray data will be used as a membership criterion, particularly important for diskless cluster members. \par
	In Sect. \ref{optdata} we will discuss the reduction of the optical data; in Sect. \ref{mastercat} we will present NIR and X-ray data and the realization of the multiband catalog; in Sect. \ref{ccd} we will use the optical Color-Magnitude diagrams to estimate distance, average extinction and age spread of the cluster; in Sect. \ref{Qsel} we will identify all the stars with circumstellar disks; in Sect. \ref{spadis} we defined the cluster members and studied the structure of NGC~6611 and in Sect. \ref{photoevaporation} we will verify the effects of the UV radiation of the massive stars on the spatial distribution of the circumstellar disks.	
	

\section{Optical data}
\label{optdata}
	 The optical data have been taken in the BVI bands with the \textit{Wield Field Camera} (WFI), mounted on the 2.2 meter telescope of the \textit{European Southern Observatory} (ESO) in La Silla (Chile). 


\subsection{Optical observations and data reduction.}

	The optical images used in this work were taken on 29 July 2000 with WFI, a mosaic of $4 \times 2$ CCDs of $2048 \times 4096$ pixels, separated by gaps of $23^{\prime \prime}.8$ in right ascension and $14^{\prime \prime}.3$ in declination, for a filling factor of 95.5\%.  The field of view (FOV) of each CCD is then $8^{\prime}.12 \times 16^{\prime}.25$, and the total WFI FOV is $33^{\prime} \times 34^{\prime}$. \par

\begin{figure}
	\centering	
	\includegraphics[width=9cm]{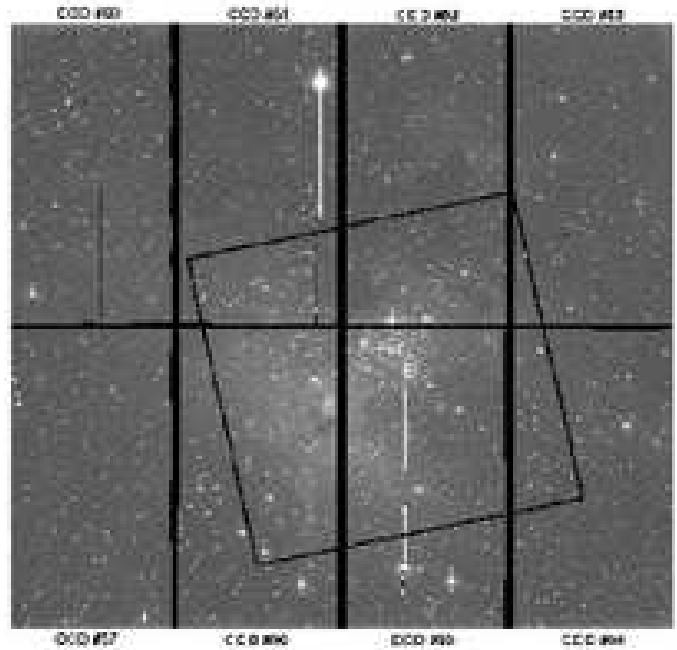}
	\caption{WFI image in band I of NGC~6611. The exposure time was 240 sec. The figure also shows the ACIS FOV, centered in the cluster, and the identifications of the CCDs of WFI.}
	\label{wfiI}
	\end{figure}

	The observations are part of the program ESO Imaging Survey (EIS), including also the PRE-FLAMES program \citep{Mom2001}. Details of the 10 images of NGC~6611 are given in Table \ref{tabellaesposizioni}: the seeing conditions were good during all the night, which was of photometric quality. Figure \ref{wfiI} shows a WFI image in I band, with the ACIS FOV (see Sect. \ref{mastercat}) also indicated. The cluster is in the center of the image. In addition 19 images of 3 standard fields were used for the photometric calibration (see Sect. \ref{calibrazione}).  \par
\par   

\begin{table*}[!ht]
\centering
\caption {WFI observations of the open cluster NGC~6611, obtained on 29 July 2000.}
\vspace{0.5cm}
\begin{tabular}{ccccc}
\hline
\hline
Image& Exposure time(sec) & Starting Time& Filter& Seeing(arcsec) \\
\hline
$NGC~6611-1$    &$30 $   &00:17:02.136        &$Ic/lwp$       &$1.17$\\
$NGC~6611-2$    &$240$   &00:19:18.375        &$Ic/lwp$       &$1.31$\\
$NGC~6611-3$    &$240$   &00:25:00.508        &$Ic/lwp$       &$1.07$\\
$NGC~6611-4$    &$30 $   &00:31:28.294        &$V/89  $       &$1.43$\\
$NGC~6611-5$    &$240$   &00:33:43.779        &$V/89  $       &$1.31$\\
$NGC~6611-6$    &$240$   &00:39:21.946        &$V/89  $       &$1.43$\\
$NGC~6611-7$    &$30 $   &00:46:09.205        &$B/99  $       &$1.43$\\
$NGC~6611-8$    &$30 $   &01:00:40.413        &$B/99  $       &$1.07$\\
$NGC~6611-9$    &$240$   &01:02:21.695        &$B/99  $       &$1.07$\\
$NGC~6611-10$   &$240$   &01:07:40.906        &$B/99  $       &$1.17$\\
\hline
\hline
\multicolumn{4}{l} {} 
\end{tabular}
\label{tabellaesposizioni}
 \end{table*}

	Instrumental calibration (\textit{Bias}, \textit{Overscan} and \textit{Flat-Field} corrections) was performed with {\tt IRAF} using the {\tt MSCRED} package. It was also necessary to remove the \textit{Fringe Pattern} from the images in I band. We used the IRAF task {\tt RMVFRINGE} and the WFI fringe pattern\footnote{available at: \textit{http://www.ls.eso.org/lasilla/sciops/2p2/E2p2M/WFI/}\par \textit{CalPlan/fringing}} that reproduces the interference figure. \par
	We used the {\tt DAOPHOT II/ALLSTAR/ALLFRAME} procedures \citep{Stet87,Stet94} to detect the point sources in the WFI images and to obtain their BVI magnitudes and their positions from the PSF fitting. For all exposures, we also obtained an aperture correction to the magnitudes with the routine {\tt DAOGROW} \citep{Stet90} which uses the \textit{growth-curve} method. \par
	
   \subsection{Photometric calibration}
   \label{calibrazione}
	
	The instrumental magnitudes of stars in the standard fields SA107, SA92 and SA112 \citep{Land92,Stet00} were obtained using {\tt DAOPHOT II/ALLSTAR}. \par
	The transformations from the instrumental magnitude system to the standard Johnson-Kron-Cousin one were computed for each CCD using the routine {\tt CCDSTD}; the transformations were then applied to the standard stars by the routine {\tt CCDAVE} and finally to all the stars by the routine {\tt TRIAL} \citep{Stet05}. \par
	Usually, the typical form of the transformations for the photometric calibration is:
	
	\begin{equation}
	\label{banalsolution}
	O = M + Z + K \times{Q} + A_{1}\times{C}
	\end{equation}
	
	where $O$ are the instrumental magnitudes of the stars; $M$ are the standard magnitudes in the Johnson-Kron-Cousin system, $Z$ is the magnitude zero-point, $Q$ is the airmass of the observations, $C$ is a suitable color; $Z$, $A_1$ and the extinction coefficient $K$ are the coefficients to be determined. \par
	In our case it was necessary to include a position polynomial in the transformations, to allow for the dependence of the observed magnitudes on the position in the WFI mosaic, which is due to the non uniform illumination of the detector \citep{Man01,Koch03,Corsi03}. Since the standard fields were not observed in a sufficiently wide interval of airmasses (from 1.148 to 1.22), it was also necessary to impose the extinction term's coefficients equal to the typical values for La Silla\footnote{available at:  \textit{http://www.ls.eso.org/lasilla/sciops/2p2/E2p2M/WFI/} \par \textit{zeropoints}}: 0.14, 0.28 and 0.09 respectively for the V, B and I band. We also included a temporal term in the transformations to take into account the time variability of the extinction. Therefore, the general form of the transformations we used was:
	
	\begin{equation}
	O = M + Z + K \times{Q} + P_{color} + A_{t} \times{T} + P_{x} + P_{y}
	\label{mysolution}
	\end{equation}
     
     	where $P_{color}$ is a polynomial which takes into account color terms of first and possibly of second order; $T$ is the time of the observations; $P_{x}$ and $P_{y}$ are polynomials, with order $\leq 3$, of the coordinates of the stars in the CCD (the $x$ axis is in the West-East direction, the $y$ axis in the North-South direction). \par
	
	\begin{figure} [!ht]		
	\centering
	\includegraphics[angle=0,width=9cm]{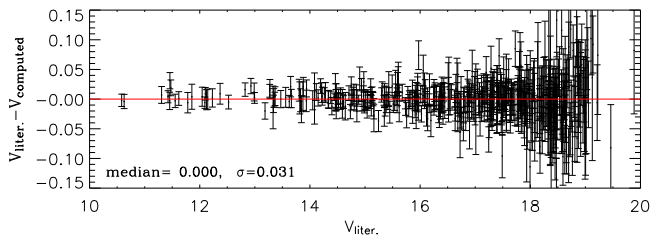}
	\caption{Difference between cataloged V magnitudes of the standard stars and those determined with the transformation (\ref{mysolution}).}
	\label{magsoluzione} 
	\end{figure}

	All the coefficients of the transformation (\ref{mysolution}) were calculated fitting the observed magnitudes (in the instrumental system) and the literature magnitudes (in the Johnson-Kron-Cousin system) of the standard stars. Fig. \ref{magsoluzione} shows the differences between the V magnitudes of the standard stars from literature and those computed with the transformation (\ref{mysolution}). The standard deviation of the distribution of the residuals is 0.031, smaller than the standard deviation (0.067) we obtained using the transformation (\ref{banalsolution}). In the other bands the results are analogous: the standard deviations obtained with the transformation (\ref{mysolution}) were 0.036 in B and 0.029 in I, systematically smaller than those derived with transformation (\ref{banalsolution}).
	
	\begin{figure} [!h]	
	\includegraphics[angle=0,width=9cm]{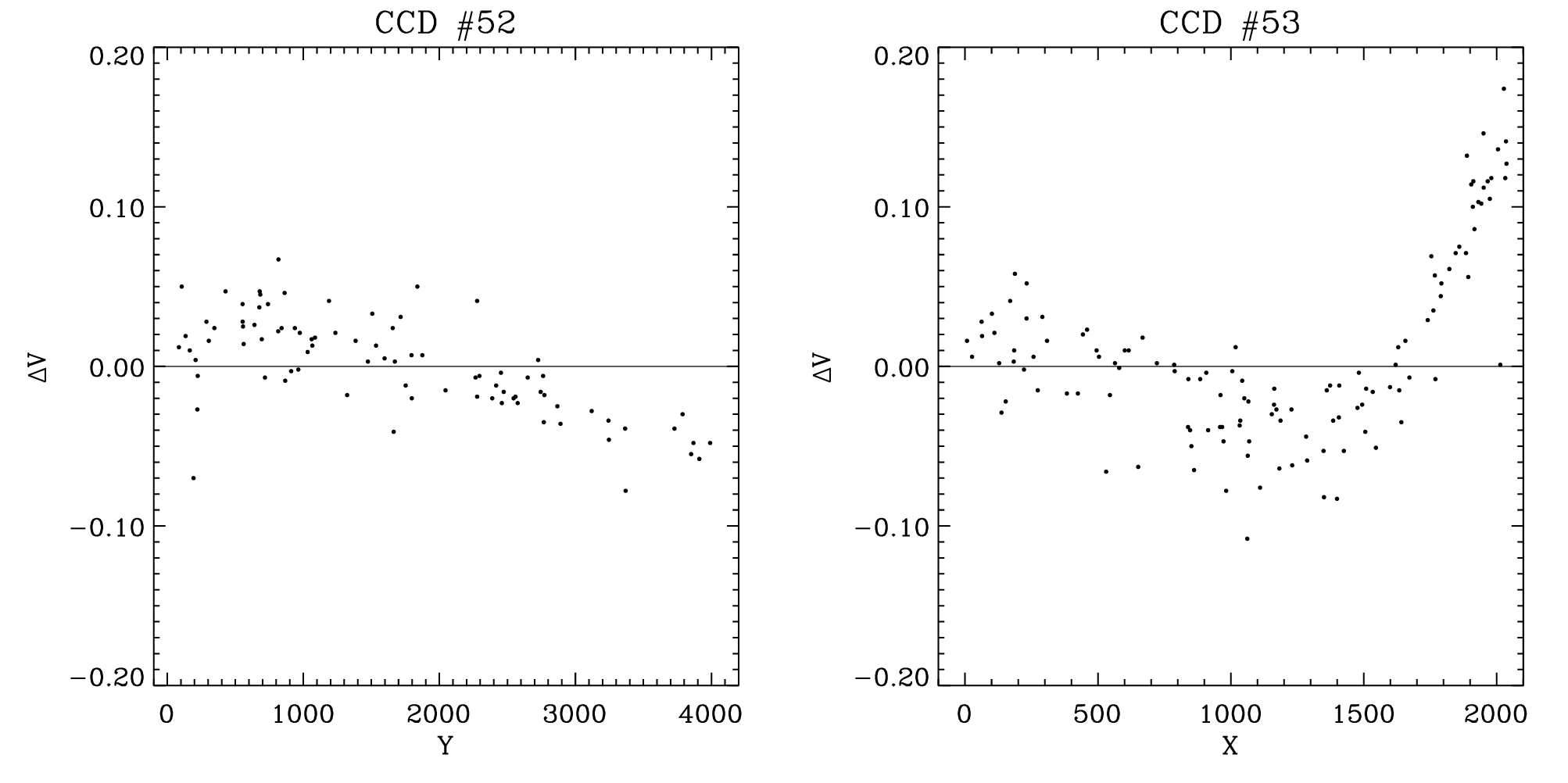}
	\includegraphics[angle=0,width=9cm]{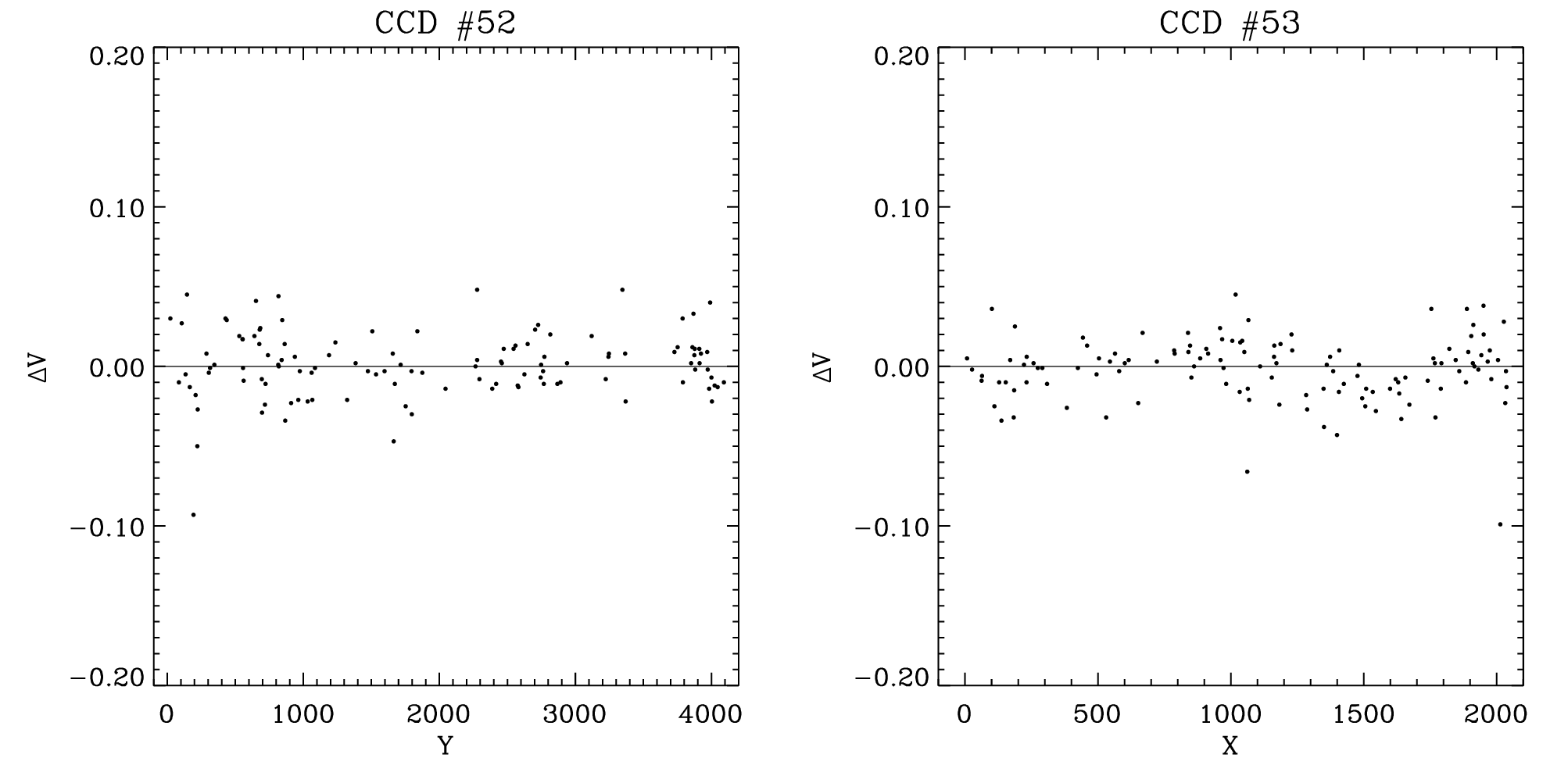}
	\caption{Residuals of the fit performed to calculate the transformations for the V magnitudes in the CCDs \#52 and \#53 vs. the spatial coordinates. In the upper panels were used two transformations without a coordinate polynomial, while in the lower panels the transformations included a coordinate polynomial.}
	\label{magpos}
	\end{figure}
	
	In order to show the need to introduce a position polynomial term in transformation (\ref{mysolution}), Fig. \ref{magpos} shows the residuals of the fit assuming no position dependence (upper panels) and with the position correction (lower panels). \par  
	For the calibration of the B magnitudes, it was necessary to add to the transformations a color-dependent extinction term $(B-V)\times Q$, with an empirically determined coefficient equal to -0.016 \citep{Stet05}, in order to account for the higher extinction in this band. \par  
	Tables \ref{coef1} and \ref{coef2} in appendix \ref{aptable} report all the terms and their coefficients used for the photometric calibration of all the CCDs.
   \subsection{Astrometry}
   \label{astrometry}
   The optical catalog obtained with the procedure described in the previous sub-sections comprised 32308 stars. We used the All-Sky Point Source Catalog from the Two-Micron All Sky Survey (2MASS) as a reference catalog to calculate the celestial coordinates of the stars from their position on the CCDs. The 2MASS catalog was chosen for its good astrometric precision (70-80 mas; \citealp{Cutri03}). First, we used the IRAF task {\tt CCXYMATCH} to find a list of common sources between the reference catalog and our optical one, with a matching tolerance of $0^{\prime \prime}.5$. In this procedure the celestial coordinates were projected in a plane with the \textit{tanx} projection (a combination of tangent plane projection and polynomials). \par
   The plate solutions were computed fitting the celestial and pixel coordinates of the common stars with the task {\tt CCMAP}. The transformations were then applied with the task {\tt SKYPIX} to obtain celestial coordinates of the stars. \par
\begin{figure} [!h]	
	\centering
	\includegraphics[angle=0,width=6cm]{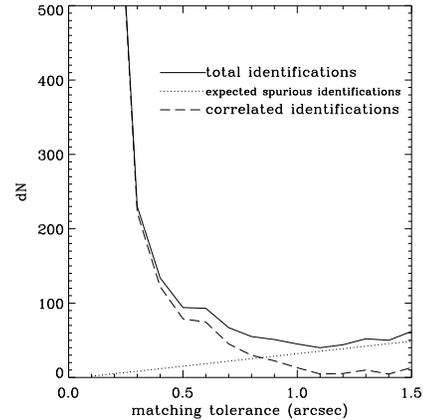}
	\caption{Differential distributions of the total, spurious and correlated detections of the stars in common between optical and 2MASS catalogs. For tolerances $>1^{\prime \prime}$ almost all the identifications are spurious.}
	\label{ident}
	\end{figure}

   The astrometric precision of the optical catalog was evaluated by matching the optical and 2MASS catalogs with different matching tolerances (ranging between $0^{\prime \prime}.1$ and $2^{\prime \prime}.0$). We used a statistical procedure that allows us to separate the correlated and spurious identifications distributions \citep{Dami06}. To find the astrometric precision, we compared the distributions of the spurious and correlated identifications in annuli with increasing radii (see Fig. \ref{ident}). We then studied the distribution of only the correlated identifications, finding that our final astrometric precision was equal to $0^{\prime \prime}.6$ at 2$\sigma$ (i.e. two truly correlated sources have a probability of $\sim$95.4\% of being identified).
   
   \subsection{Comparison with previous catalogs}
   \label{conf}
	\begin{figure*} []	
	\includegraphics[angle=0,width=14cm]{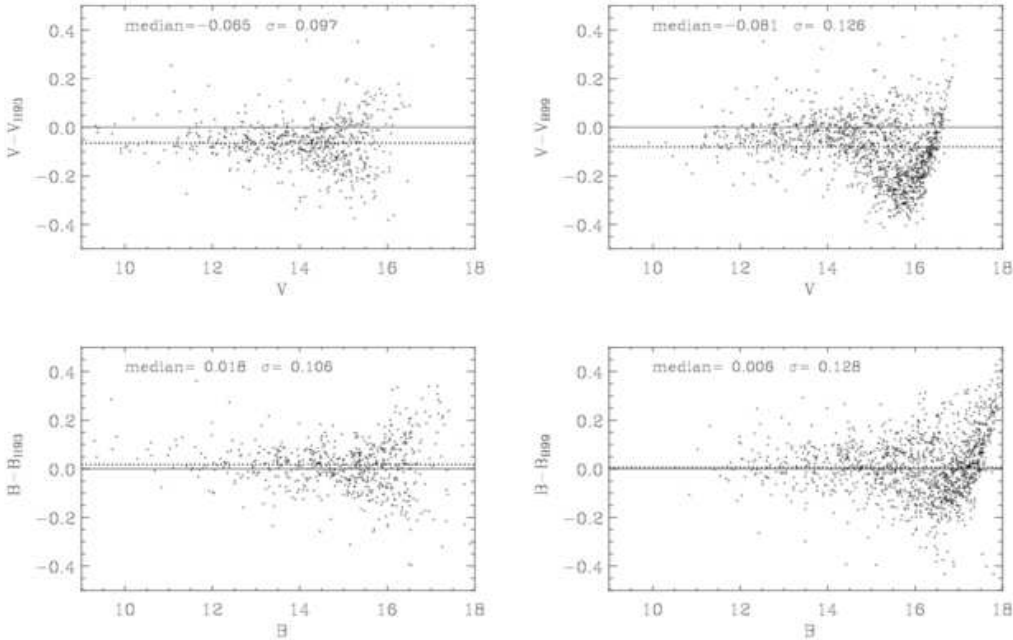}
	\caption{Differences of the magnitudes in the V (upper panels) and B (lower panels) bands of the stars common to our optical catalog and the ones of H93 (left panels) and B99 (right panels). The dotted horizontal line marks the median magnitude difference between the two samples.}
	\label{confV}
	\end{figure*}

   We compared our optical catalog with the two previous catalogs of the region of NGC~6611 obtained by H93 and B99. The first, consisting of a total of 1026 stars with B, V, J, H and K photometry and limited to V$\leq 17^m$, was based on observations of the Kitt Peak National Observatory (KPNO); the latter is a compilation of previous catalogs including the one of H93, with suitable photometric transformations, and consisting of 2185 stars with V$\leq 17^m$. \par
   To match these catalogs with ours, we first used a matching tolerance of $2^{\prime \prime}$, finding systematic differences in celestial positions between the common stars: $-0^{\prime \prime}.870$ in right ascension and $+0^{\prime \prime}.177$ in declination for the catalog of H93, and $-0^{\prime \prime}.263$ in right ascension and $-0^{\prime \prime}.674$ in declination for the catalog of B99. We then used these values to correct the offsets. Comparing the distributions of correlated and spurious detections, we estimated that the best matching tolerances were equal to $0^{\prime \prime}.3$ and $0^{\prime \prime}.5$ for the catalog of H93 and B99 respectively. \par
   We then calculated the differences in magnitude of the common stars, with the results reported in Fig. \ref{confV}. Comparing the median values of these differences (-0.065 and -0.081 in V band, 0.018 and 0.006 in B band) with the limit for acceptable systematic differences (taking into account anomalous extinction, flat-fielding errors, ecc..) of $\sim 0.02^m$ \citep{Stet05}, there is not a good agreement in V band. As already seen, however, the magnitudes of the standard stars are well reproduced by the photometric transformations used for the calibration and we will see later (Sect. \ref{parameter}) that all the optical Color-Magnitude diagrams achieved in this work are self consistent. \par
 
\begin{figure} [t]	
	\includegraphics[angle=0,width=9cm]{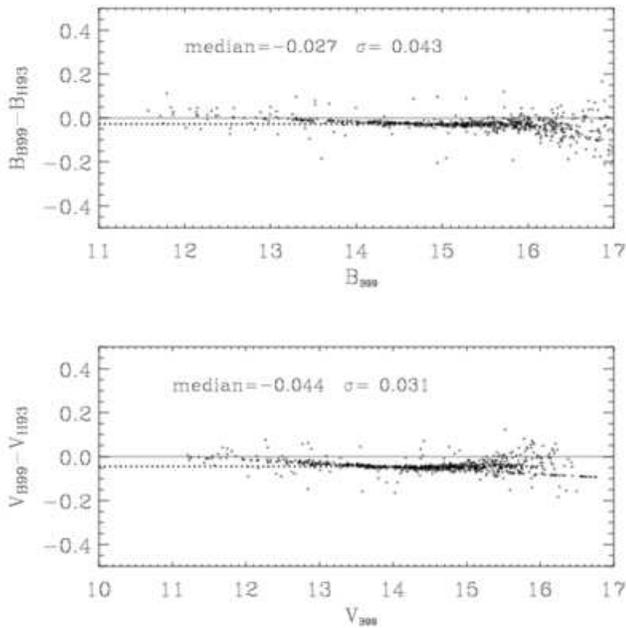}
	\caption{Differences of the magnitudes in the B and V bands of the stars common to the catalog of H93 and B99. The dotted horizontal line marks the median magnitude shift between the two samples.}
	\label{confBH}
	\end{figure}

	Fig. \ref{confBH} shows the differences of the V magnitudes for the common stars between the catalogs of H93 and B99. There is not a good agreement in the V band between these two catalogs. We ignored the differences in V magnitude between our catalog and the ones of H93 and B99, and considered our V magnitude to be the most reliable.
   
 \section{Near-infrared, X-ray data and cross-identifications}  
 \label{mastercat}
 	There are 30703 2MASS sources in the WFI FOV. Among them, we selected the NIR sources unaffected by strong photometric uncertainties. We chose to discard the sources whose magnitudes were affected by instrumental artifacts, the ones with problems in the interpolation between the stellar profile and the PSF model or in the determination of the photometric errors and the sources that are not resolved in a consistent fashion in all the bands\footnote{\textit{CC-flag} equal to $P$, $D$, $S$ and $B$; \textit{PH-qual} equal to $E$, $F$ and $X$ or \textit{PH-qual} equal to $U$ only if the \textit{rdflag} was equal to 6 \citep{Cutri03}.}. The final number of 2MASS sources selected was then 25920. \par
\begin{figure} [!h]	
	\includegraphics[angle=0,width=9cm]{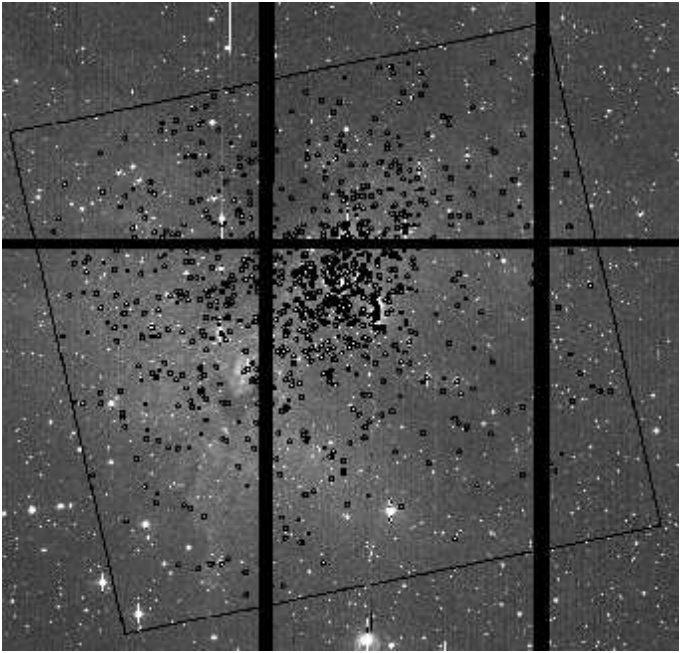}
	\caption{Detail of the WFI image of Fig. \ref{wfiI} with the ACIS-I FOV (rotated square) and the 997 X-ray sources detected (north is up and east is on the left). The X-ray sources are crowded in the center of the cluster and in the North-East quadrant.}
	\label{xsources}
	\end{figure}

   The X-ray data are taken from an archival 78 Ksec observation with \textit{Chandra} ACIS-I detector, with a FOV of $16^{\prime}.9\times16^{\prime}.9$ \citep{Lin00}. Source detection was performed with \textit{PWDetect}\footnote{available at: \textit{http://www.astropa.unipa.it/progetti\_ricerca/PWDetect}}, a wavelet-based source detection algorithm \citep{Dami97}. With PWDetect we found 997 X-ray sources in the ACIS FOV, using a 1keV exposure map and a threshold limit of 5.1 $\sigma_{sky}$, corresponding to one spurious detection on average. Fig. \ref{xsources} shows the X-ray sources of NGC~6611 overlaid on the WFI image corresponding to the ACIS FOV. This figure clearly shows that there is a crowded cluster of X-ray sources in the central region of NGC~6611.\par
   
\begin{table*} 
\footnotesize
\centering
\caption {Results of the cross-identifications.}
\vspace{0.5cm}
\begin{tabular}{rrrr}
\hline
\hline
Number of stars& Optical detection & NIR detection & X-ray detection \\
\hline
$12561$		&$yes$		&$no$		&$no$\\
$16390$		&$no$		&$yes$		&$no$\\
$9156$		&$yes$		&$yes$		&$no$\\
$332$		&$yes$		&$no$		&$yes$\\
$33$		&$no$		&$yes$		&$yes$\\
$522$		&$yes$		&$yes$		&$yes$\\
$38995$         &\multicolumn{3}{r} {Total of stars in the multiband catalog}\\ 
\hline
\hline
\multicolumn{4}{l} {The X-ray sources without optical and 2MASS counterpart are 148.} 
\end{tabular}
\label{mastercatalogo}
 \end{table*}	 
  
  In order to obtain the multiband catalog of NGC~6611, we matched the common sources between the optical, near-infrared and X-ray catalogs. First, we matched near-infrared and optical detections, with a matching tolerance of $0^{\prime \prime}.7$, slightly larger than the 2$\sigma$ astrometric precision of the optical catalog ($0^{\prime \prime}.6$, see Sect. \ref{astrometry}). The differential distributions of correlated and spurious identifications (Fig. \ref{ident}) show that at the chosen threshold of $0^{\prime \prime}.7$ the correlated identifications are about twice the spurious ones. We have discarded the optical sources detected only in one band and without a NIR counterpart. The results of all the cross-identifications are reported in Table \ref{mastercatalogo}: with this choice and the tolerance of $0^{\prime \prime}.7$ we found 9680 (522+9156 in Table \ref{mastercatalogo}) optical-NIR sources, with 355 multiple identifications, 12893 (12561+332) optical sources without a near-infrared counterpart and 16423 (16390+33) infrared sources without an~optical counterpart.\par
	Then, we matched all the optical or NIR sources with the X-ray sources. We could not use a constant matching tolerance for all X-ray sources because the PSF in the ACIS-I images is not uniform: the FWHM is $\sim0^{\prime \prime}.5$ on-axis and becomes much larger near the edges of the field. We used a tolerance proportional to the position uncertainties derived by PWDetect (equal to $n\times \sigma_x$, where $\sigma_x$ is the uncertainty). To obtain the distribution of the spurious detections we performed 40 cross-identifications with different values for $n$ and with a set of different rigid translations of the X-ray sources coordinates. We set the value of $n$ at 1.5 comparing the distribution of spurious and total identifications. With this matching tolerance we found 887 X-ray sources with optical or near-infrared counterparts, including 78 multiple identifications. Table \ref{cat} shows a portion of the multiband catalog: the first column reports the ID of the stars; the second and third their celestial coordinates; the columns from 4$^{th}$ to 15$^{th}$ report the optical and NIR magnitudes with their uncertainties; the 16$^{th}$ is a flag indicating if the star has an X-ray counterpart; the 17$^{th}$ and 18$^{th}$ columns are the $\chi^2$, calculated in the optical PSF fitting procedure, and {\tt SHARP} values, a parameter calculated in the source detection procedure and that takes into account the sharpness of the source, allowing one to discard the extended sources and the events due to cosmic rays \citep{Stet87}; the 19$^{th}$ column is a flag indicating if the source has a multiple identification between the catalogs matched to obtain the multiband catalog.	 \par	 
       
\begin{sidewaystable*}
\begin{minipage}[t][180mm]{\textwidth}
\caption{Part of the electronic multiband catalog.}\label{cat}
\centering
\begin{tabular}{ccccccccccccccccccc} 
\hline\hline  
ID&RA(J2000)&DEC(J2000)&$B$&$\sigma_B$&$V$&$\sigma_V$&$I$&$\sigma_I$&$J$&$\sigma_J$&$H$&$\sigma_H$&$K$&$\sigma_K$&X-ray emission\footnote{If equal to Y then the star has an X-ray counterpart.}&$\chi^2$&{\tt SHARP}&mul\footnote{If equal to 1 the star is characterized by a multiple identification.}\\
\hline
\hline 
   1760& 274.847168& -13.750009& 23.40& 0.14& 21.26& 0.04& 18.91& 0.04&  NA  &  NA  &  NA  &  NA   &  NA  &  NA       &  N  	 & 1.392&-0.028&    0\\
   1761& 274.730042& -13.749800& 22.46& 0.03& 20.10& 0.04& 17.33& 0.01& 15.42& 0.05& 14.27& 0.06& 13.74& 0.08&  N   & 1.563& 0.000&    0\\
   1762& 274.753571& -13.749639& 22.60& 0.04& 20.40& 0.02& 17.53& 0.02& 15.54& 0.06& 14.58& 0.06& 13.53&  NA  &Y& 1.811& 0.601&    1\\
   1763& 274.733673& -13.749538& 21.98& 0.01& 19.79& 0.02& 16.93& 0.01&  NA  &  NA  &  NA  &  NA  &  NA   &  NA  & Y& 1.563& 0.119&    0\\
   1764& 274.727478& -13.749289& 23.21& 0.10& 20.91& 0.04& 17.90& 0.01& 16.25& 0.11& 14.04&  NA  & 12.60&  NA      &  N      & 1.546& 0.059&    0\\
   1765& 274.752350& -13.749180&  NA  &  NA & 23.03& 0.19& 19.75& 0.07&  NA  &  NA  &  NA  &  NA  &  NA  &  NA     &  N      & 1.902&-1.558&    0\\
   1766& 274.828247& -13.749168& 22.88& 0.05& 21.09& 0.03& 18.74& 0.02&  NA  &  NA  &  NA  &  NA  &  NA  &  NA     &  N       & 1.381& 0.203&    0\\
   1767& 274.731903& -13.748953& 21.78& 0.05& 19.48& 0.01& 16.80& 0.01& 14.83& 0.11& 13.68& 0.09& 13.05& 0.06& Y& 1.573&-0.091&    0\\
   1768& 274.714508& -13.748638&  NA  &  NA &  NA  &  NA &  NA  &  NA & 16.12& 0.12& 13.77& 0.06& 12.67& 0.04& N 	  & 0.000& 0.000&    0\\
   1769& 274.827972& -13.748659&  NA  &  NA &  NA  &  NA  & NA  &  NA & 16.01& 0.10& 14.65& 0.06& 13.97& 0.06&  N 	  & 0.000& 0.000&    0\\
   1770& 274.737183& -13.748060& 15.56& 0.01& 14.66& 0.01& 13.60& 0.01& 12.70& 0.02& 12.35& 0.02& 12.12& 0.02& Y& 1.695& 0.049&    0\\
   1771& 274.738922& -13.747979& 23.97& 0.26& 22.11& 0.10& 18.67& 0.02& 16.44& 0.13& 15.13&  NA  & 14.30&  NA  &  N        & 1.609& 0.255&    0\\
   1772& 274.804474& -13.747979& 24.32& 0.23& 22.31& 0.07& 19.55& 0.04&  NA  &  NA  &  NA  &  NA  &  NA  &  NA  &  N       & 1.534& 0.491&    0\\
   1773& 274.817169& -13.747875& 20.79& 0.01& 19.32& 0.01& 17.60& 0.01& 16.15& 0.11& 15.27& 0.11& 14.25&  NA  &  N         & 1.162& 0.033&    0\\ 
   1774& 274.829285& -13.747687& 24.58& 0.20& 22.56& 0.30& 19.63& 0.11&  NA  &  NA  &  NA  &  NA  &  NA  &  NA  &  N       & 1.647& 0.013&    0\\
   1775& 274.726166& -13.747549& 20.17& 0.01& 18.12& 0.01& 15.63& 0.01&  NA  &  NA  &  NA  &  NA  &  NA  &  NA  &  N       & 1.651& 0.073&    0\\
   1776& 274.722961& -13.747337& 23.17& 0.11& 20.97& 0.03& 18.75& 0.04&  NA  &  NA  &  NA  &  NA  &  NA  &  NA  &  N       & 1.499& 0.028&    0\\
   1777& 274.759338& -13.747359& 23.16& 0.10& 21.39& 0.04& 18.95& 0.07&  NA  &  NA  &  NA  &  NA  &  NA  &  NA  &  N       & 1.485& 0.183&    0\\
   1778& 274.756104& -13.747284& 22.18& 0.07& 20.35& 0.01& 18.26& 0.03&  NA  &  NA  &  NA  &  NA  &  NA  &  NA  &  N       & 1.311& 0.057&    0\\
   1779& 274.720978& -13.747194&  NA  &  NA & 22.64& 0.16& 18.94& 0.03&  NA  &  NA  &  NA  &  NA  &  NA  &  NA  &  N       & 1.816& -0.549&	0\\
   1780& 274.710754& -13.747069& 23.03& 0.27& 21.00& 0.05& 17.75& 0.03&  NA  &  NA  &  NA  &  NA  &  NA  &  NA  & Y& 1.601&  0.010&   0\\
   1781& 274.806885& -13.747075& 22.61& 0.10& 20.46& 0.02& 17.45& 0.01& 15.37& 0.06& 14.37& 0.05& 14.02& 0.08&  N    	    & 1.630&  0.145&   0\\
   1782& 274.764679& -13.746936& 24.01& 0.29& 22.23& 0.14& 19.10& 0.08&  NA  &  NA  &  NA  &  NA  &  NA  &  NA  &  N 	    & 1.530&  0.050&   0\\
   1783& 274.707275& -13.746615&  NA  &  NA & 21.92& 0.37& 18.50& 0.04& 14.96& 0.05& 13.40& 0.05& 12.43& 0.04&  N    	    & 1.712& -0.387&   0\\
   1784& 274.804321& -13.746662&  NA  &  NA & 22.59& 0.25& 20.08& 0.07&  NA  &  NA  &  NA  &  NA  &  NA  &  NA  &  N 	    & 1.544& -0.059&   0\\
   1785& 274.746338& -13.746284& 23.35& 0.20& 20.98& 0.03& 17.88& 0.01& 15.61& 0.07& 14.55& 0.05& 13.99& 0.06&  N    	    & 1.510&  0.156&   0\\
   1786& 274.791351& -13.488843&  NA  &  NA & 21.92& 0.11& 18.65& 0.05&  NA  &  NA  &  NA  &  NA  &  NA  &  NA  &  N 	    & 2.398& -0.308&   0\\
   1787& 274.801392& -13.745899& 23.08& 0.08& 20.70& 0.02& 17.75& 0.01& 15.46& 0.05& 14.31& 0.03& 13.48& 0.03&  N    	    & 1.903&  0.086&   0\\
   1788& 274.726746& -13.745795& 24.59& 0.38& 22.33& 0.09& 19.48& 0.04&  NA  &  NA  &  NA  &  NA  &  NA  &  NA  &  N 	    & 1.612&  0.256&   0\\
\hline
\multicolumn{19}{l}{If $\chi^2$ and {\tt SHARP} are equal to 0 then the source was observed in only an optical band, and was not possible to calibrate the magnitude.} \\ 
\multicolumn{19}{l}{NA is for "Not Available".}
\end{tabular}
\vfill
\end{minipage}
\end{sidewaystable*}

\section{The color-color and color-magnitude diagrams.}
\label{ccd}
	
	 Using the combined X-ray and optical data of the catalog produced in this work, extended to much fainter stars (V$\leq 23^m$) than previous works, we estimated new values for distance, age spread and average extinction of NGC~6611.  \par

\subsection{Distance of NGC~6611}	 	

	\begin{figure} [!]	
	\includegraphics[angle=0,width=9cm]{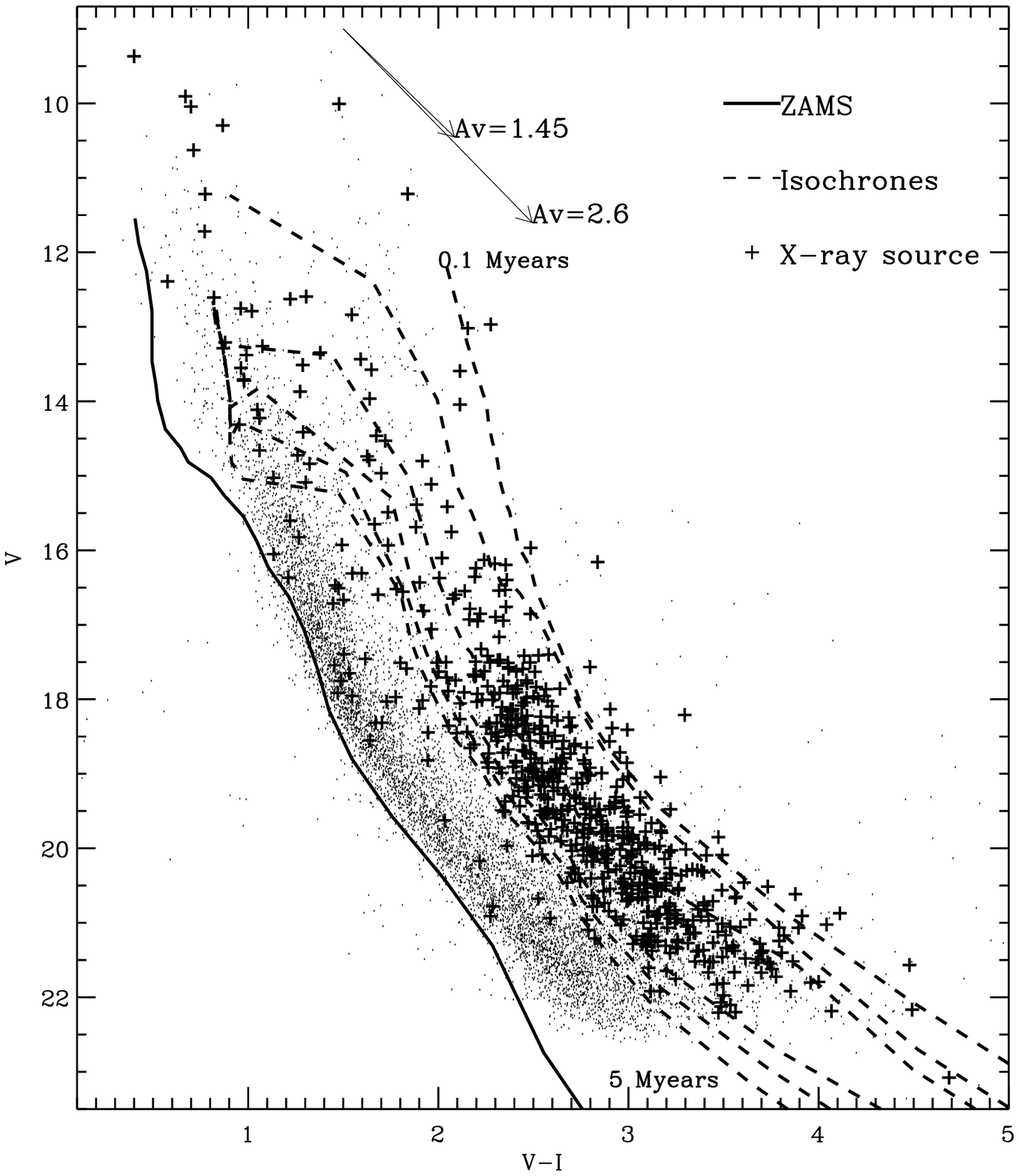}
	\caption{$V$ vs $V-I$ diagram of the stars in the WFI FOV. Crosses mark the X-ray sources and identify the cluster locus. The thick solid line is the ZAMS from \citet{Sies00} at the distance estimated for the cluster and with the average extinction that we have derived for the foreground stars (A$_V=1.45^m$). The dashed lines are the isochrones at 0.1, 0.25, 1, 2.5, 3 and 5 million years, with the average extinction derived for the cluster members (A$_V=2.6^m$). The extinction vectors are obtained from the relations of \citet{Muna96} and are drawn both for the canonical reddening law and for the one of the cluster (see Sect. \ref{parameter}).}
	\label{VVI}
	\end{figure}
	 
    Fig. \ref{VVI} shows the $V$ vs. $V-I$ diagram of those stars in the WFI FOV having errors in $V$ and $V-I$ smaller than 0.1$^m$. To estimate the distance of NGC~6611 we took advantage of the fact that the cluster is within the Eagle Nebula. Because of the presence of dust associated with the nebula, the cluster stars are more extinguished than the foreground stars, not obscured by the cloud. The background main sequence stars are even more reddened, so in the diagram in Fig. \ref{VVI} they are not visible or are shifted in the direction of the reddening vector. For this reason the main sequence stars up to the distance of the cloud in the diagram in Fig. \ref{VVI} form a locus limited by the ZAMS (here and in the following we use the Siess isochrones; \citealt{Sies00}). We then estimated the distance of the cluster and the average extinction of the foreground stars fitting the ZAMS to this locus, as in the study of \citet{Prisi05} of the cluster NGC~6530. We used the extinctions laws for the optical bands of \citet{Muna96}, with R$_V$=3.1. \par
    We found that the distance of the cluster is about 1750~pc and the average extinction of the field stars is $A_V = 1.45^m$. The distance of NGC~6611 we estimated is then slightly smaller than the values of H93 (obtained from the distance modulus of 36 OB stars) and of B99 (obtained with 72 probable cluster members, using a membership criterion based on the proper motions), while it is in good agreement with the value of \citet{Bona06}, obtained with 2MASS data, and with the value reported in the Catalog of Galactic Open Clusters of \citet{Khar05}. The difference between the distance values found in this work and in H93 is compatible with the difference in the V magnitude between the two catalogs. Indeed we obtained a distance of about 2 kpc with the procedure described above after correcting the V magnitudes for the median value of the differences shown in Fig. \ref{confV}. \par 

\subsection{Reddening law and age spread.}
\label{parameter}	 	

	It is well known that young stars are characterized by an intense X-ray emission. We can assume that the large majority of the 997 X-ray sources detected in the ACIS observation are young stars, members of NGC~6611, as in the study of \citet{Damiani04} of the similar young cluster NGC~6530.\par
	All previous works suggest that the reddening law toward NGC~6611 is slightly different to the canonical law (R$_V=3.1$). H93 found R$_V=3.75$, and this value was used in most later works. To determine R$_V$, H93 calculated the Johnson index $Q$ for all the stars in their catalog with emission in K. Then they used the relations \citep{Jon66}:
      
      \begin{equation}
      (B-V)_0=0.332 \times Q $\hspace{1.5cm}$
      (V-K)_0=1.05 \times Q
      \end{equation}     
   
	which hold for main-sequence stars with negative Q (spectral class A0 or earlier); then they obtained the value of R$_V$ from the equation of \citet{Stee91}:
	
	\begin{equation}
	R_V=1.1 \times{E_{V-K}/E_{B-V}} 
	\end{equation}

	\begin{figure} []	
	\includegraphics[angle=0,width=9cm]{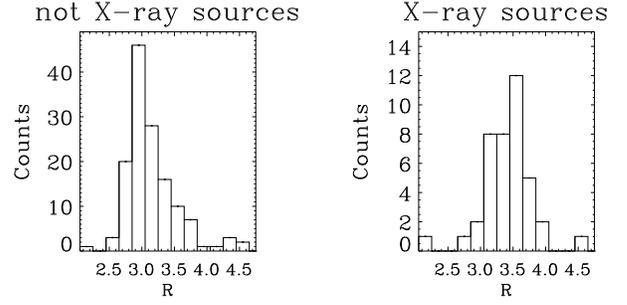}
	\caption{Histograms of R$_V$ calculated with the U magnitudes and the procedure (described in the text) of H93 and our B, V and K magnitudes. The left panel shows the distribution of the stars not detected in the X-ray observation, while the right panel that of the stars detected. In all the panels we used all the stars common to our catalog and to H93 and with negative Q.}
	\label{Rhisto}
	\end{figure}

	and found a bimodal distribution of R$_V$: one peak is at the canonical value ($R_V=3.1$, field stars) and the other at $R_V=3.75$ (cluster members). It has been shown (see Sect. \ref{conf}) that the V magnitudes obtained in our work are not consistent with those of H93. The K magnitudes are not in good agreement either, with systematic differences as for the V magnitudes. We chose then to repeat the procedure of H93, using our B, V and K magnitudes and the U magnitudes of H93. Fig. \ref{Rhisto} shows two histograms of R$_V$, obtained for the stars without and with X-ray emission. The left panel clearly shows that the peak of the distribution of no X-ray sources is at R$_V \sim3$, while for the X-ray emitting stars, largely dominated by cluster members, the peak is at $R_V\sim3.5$, with a median value of $R_V=3.27$, that we adopted in the following (right panel). This confirms the anomalous reddening law toward NGC~6611, even if less extreme than in H93. \par 
	Once having chosen the reddening law for the cluster members, it was possible to redden the theoretical isochrones, from \citet{Sies00}, in the optical Color-Magnitude diagrams and to estimate the age of the X-ray sources. In most of the previous works the age of PMS population of NGC~6611 was estimated to be smaller than 1 Myr. For example, H93 found that the PMS stars with masses M $\geq 3M_{\odot}$ have an age ranging from 0.25 to, at least, 1 Myr. As shown in Fig. \ref{VVI}, our data suggest that pre main-sequence stars of NGC~6611 with masses smaller than 1 $M_{\odot}$ (with a V magnitude greater than V=$17.5^m$) may have an age in 0.1-3 Myr range, with a median value of 1 Myr. \par 
	A rough value for the average extinction for the stars of NGC~6611 (A$_V=2.6^m$) was estimated by fitting the bright part of the isochrones older than 1 million years and of the ZAMS from \citet{Keha95} to the bright stars with X-ray emission. The cluster parameters (distance, extinction and range of age) derived here from a $V$ vs. $V-I$ diagram are consistent with the parameter obtainable from the other optical Color-Magnitude diagrams.\par


	\begin{figure*} []	
	\includegraphics[angle=0,width=16cm]{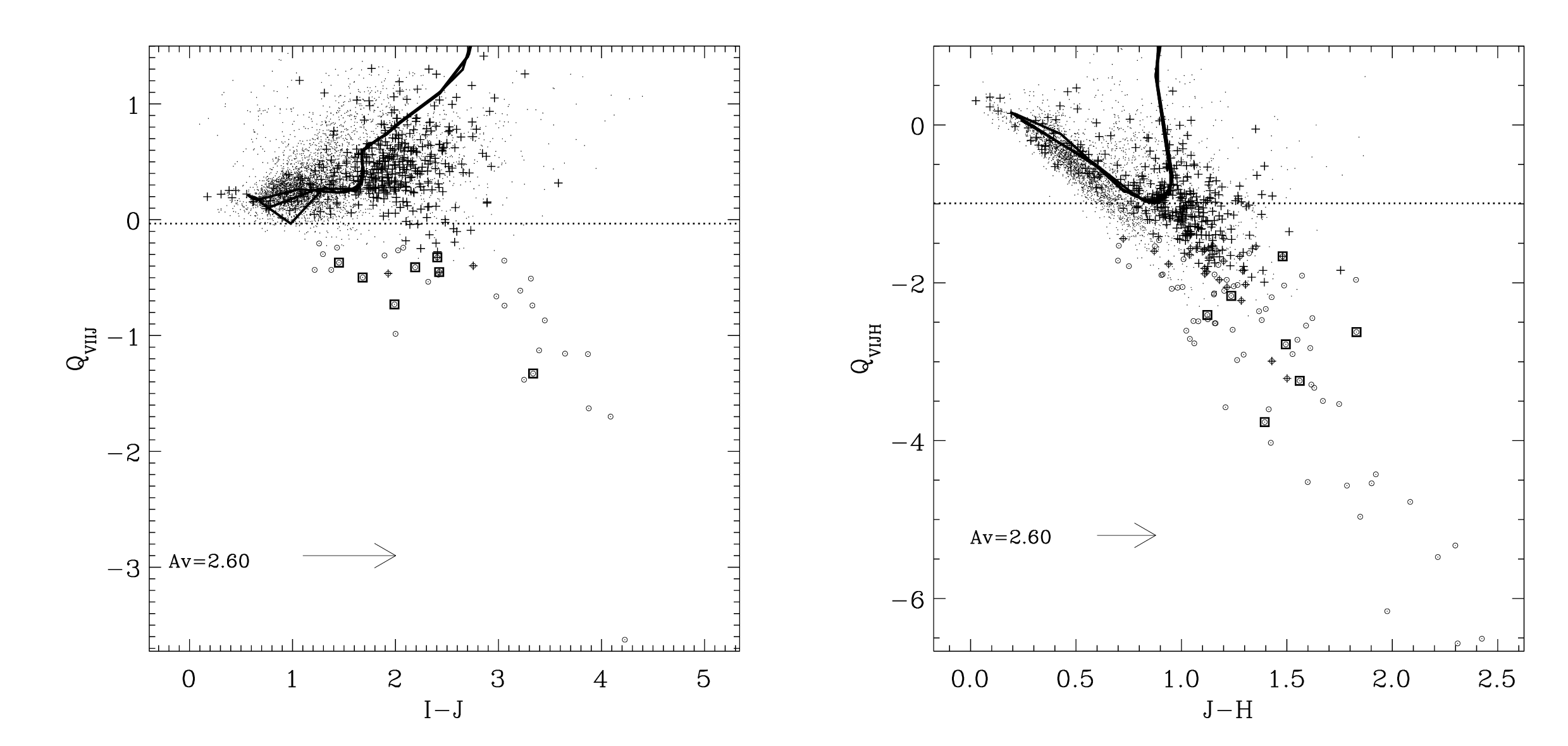}\par
	\includegraphics[angle=0,width=16cm]{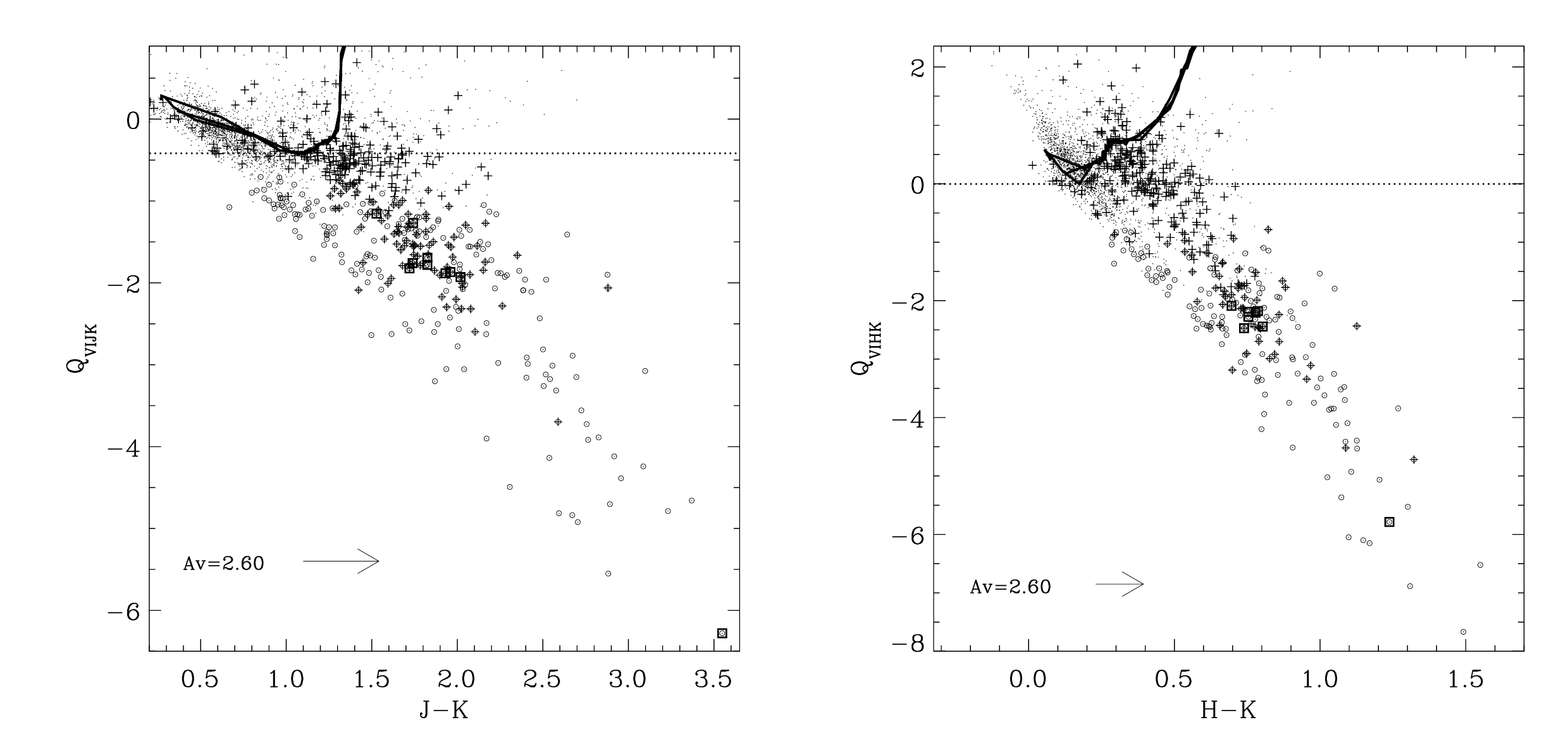}\par
\hspace{0.2cm}	\includegraphics[angle=0,width=7cm]{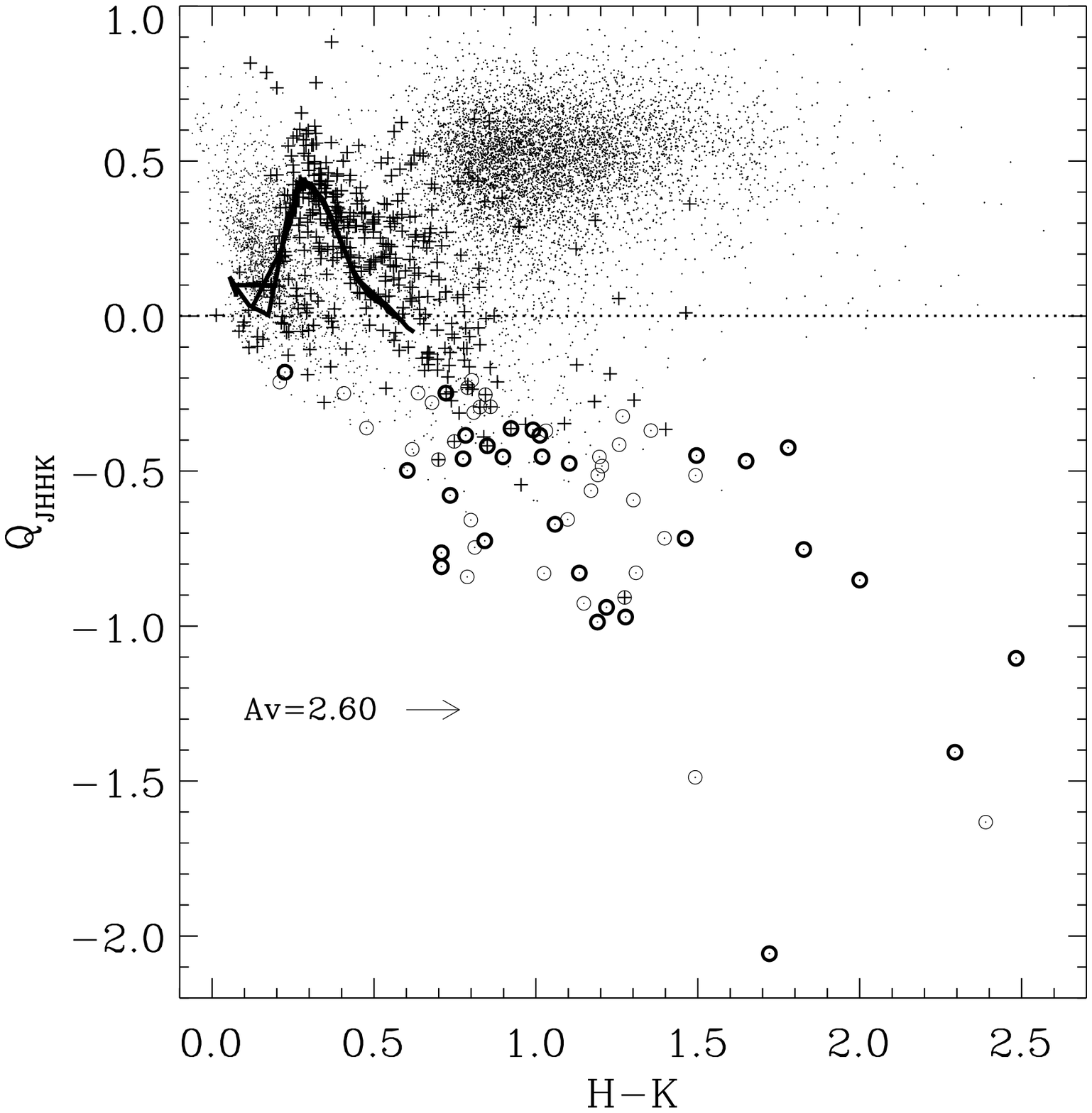}
	\caption{Diagrams of the extinction-independent indices vs. the various photometric colors, used for the selection of the stars with NIR excess in the WFI FOV. The circles identify those stars with indices significantly smaller than the limit for the photospheric emission (dashed lines); the X-ray sources are represented by $+$, while the stars with multiple identifications with small squares. The thick solid lines are the photospheric stellar colors. In the $Q_{JHHK}$ (lowest left) diagram the thicker circles mark the stars with excess but without optical emission, that are candidates embedded protostars.}
	\label{Qgraf}
	\end{figure*}

\section{Stars with circumstellar disks}
\label{Qsel}
	
	We have selected the stars with circumstellar disks by searching for infrared excesses in their emission. To do this, we used the extinction-independent indices analogous to the Johnson index, as defined in \citet{Dami06}. The general definition of these indices is: 
	
	\begin{equation}
	\label{Qeq}
	Q_{ABCD}=(A-B)-(C-D) \times{E_{A-B}/E_{C-D}}
	\end{equation}
	
	where, $A$, $B$, $C$ and $D$ are magnitudes in four bands and $E_{A-B}$ and $E_{C-D}$ are the corresponding color excesses. The ratios between the reddening coefficients was obtained from the extinction laws of \citet{Muna96}, for the optical bands, and from \citet{Rile85} for the NIR colors. We chose these laws by comparing the extinction vectors with the slope of the locus of the background stars in the infrared Color-Color diagrams.\par
	We made the assumption that the $V-I$ color is a characteristic of the photospheric emission of the stars. This assumption may be not strictly valid for stars characterized by strong accretion \citep{Harti90,KH90}. We then defined five indices to reveal excess emission in the NIR bands: $Q_{VIIJ}$, $Q_{VIJH}$, $Q_{VIJK}$, $Q_{VIHK}$, $Q_{JHHK}$. \par 
		
	Fig. \ref{Qgraf} shows the diagrams used for the selections of the stars with infrared excess. To be conservative, we performed the selection only for stars with errors, in the considered colors, smaller than $0.15^m$. Since the $Q$ indices are independent of the extinction, the extinction vectors are parallel to the $x$-axis in Fig. \ref{Qgraf}. We compared the indices of the stars with a lower limits for the photospheric emission, chosen from the normal-stars colors by \citet{Keha95}. Only stars with an index smaller than the limit by more than 3 times the $\sigma_Q$ are considered to show NIR excess. In order to account for spurious cross-matches between infrared and optical catalogs that would affect the selection of the stars with infrared excess, we calculated the $Q$ indices only for the optical-NIR stars identified within a matching radius of $0^{\prime \prime}.3.$ \par
	
	\begin{table} 
	\centering
	\caption {Number of stars in WFI and ACIS FOV and the number of X-ray sources with excesses in the infrared bands specified in the first column.}
	\vspace{0.5cm}
	\begin{tabular}{cccc}
	\hline
	\hline
	Bands        & WFI FOV        & ACIS FOV     &  $X-ray$\\
	\hline
	$J$            &$12 $	       &$5  $	       &$0 $\\
	$H$ 	       &$7  $	       &$1  $	       &$0 $\\
	$K$ 	       &$234$	       &$138$	       &$71$\\
	$J+H$          &$0  $	       &$0  $	       &$0 $\\
	$J+K$	       &$5  $	       &$4  $	       &$1 $\\
	$H+K$	       &$67 $	       &$40 $	       &$19$\\
	$J+H+K$	       &$17 $	       &$9  $	       &$3 $\\
	\hline
	$Total$	       &$342$          &$197$          &$94$\\	
	\hline
	\hline
	\end{tabular}
	\label{NQ}
	\end{table}	
	
	Table \ref{NQ} summarizes the number of stars and X-ray sources (N$_{X-ray}$) with excess in the NIR bands indicated in the first column. There are very few stars with excess \textit{only} in the J or H bands (first and second rows), and none with excess in only these two bands together (fourth row). The excess in K band is more typical of the presence of an inner disk. The nature of the sources with excess only in J and H might be explained with additional data at longer wavelength. Also only $\sim$50\% of the stars with a circumstellar disk are X-ray sources. The selection of young stars with the X-ray emission may be incomplete when applied to a star formation region, with a large number of stars with a circumstellar disk. The lack of X-ray emission in a substantial fraction of stars with disk is consistent with the higher L$_X$ of WTTS than CTTS observed in other star forming regions, such as in Orion \citep{Prei05,Flacco03,Flacco2003}.\par 	

\section{Membership and cluster structure}
\label{spadis}

	Having detected the stars with a circumstellar disk, we are able to identify the cluster members. The ACIS observation detected 997 X-ray sources (see Sect. \ref{mastercat}), mostly members of the cluster. 148 of the X-ray sources have neither an optical nor 2MASS counterpart (see Table \ref{mastercatalogo}). We do not consider these stars as cluster members, even if their large number and their clustering may suggest that at least some of them are members embedded in the cloud. We have also excluded 18 X-ray sources without a circumstellar disk, because in the $V$ vs. $V-I$ optical diagram they have a position compatible with young foreground MS stars. Adding to this sample the stars with a circumstellar disk that are not X-ray sources, we found 1122 cluster members in the WFI FOV (990 in the ACIS FOV). \par
	The number of cluster members found in this work is larger than in the previous works. H93 found a total of 174 cluster members, using a criterion that allowed them to consider only stars with spectral type earlier than A0. In Fig. \ref{VVI} the V magnitude of the stars with spectral class earlier than A0 is brighter than 15$^m$, and X-ray sources are almost all fainter than this limit. Using a membership probability distribution based on the stars' projected position and proper motion, B99 found 87 highly probable and 525 probable members of NGC~6611. In addition to the fact that our catalog is deeper than the one of B99 (see Sect. \ref{conf}) the accuracy of a proper motion membership criterion is questionable at the large distance of NGC~6611. \par
	With the sample of cluster members we defined, we studied the radial density profile of the cluster. We used the 2-parameter density profile of \citet{King66}: 
	
	\begin{equation}
	\sigma(r)=\frac{\sigma_0}{1+\left(r/r_{core}\right)^2}
	\label{denpro}
	\end{equation}

	\begin{figure} []	
	\includegraphics[angle=0,width=8cm]{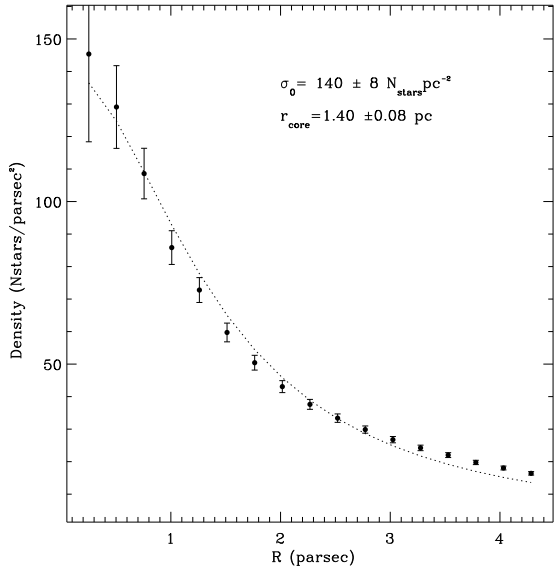}
	\caption{Radial density profile found for the cluster members. The points are the observed density in circles with the radii shown in abscissa; the dotted line is the best-fit 2-parameters King's profile (equation \ref{denpro}). The best-fit parameters are also indicated. The error bars are poissonian uncertainties.}
	\label{rdp}
	\end{figure}
	
	where $\sigma_0$ is the central cluster density, $r$ is the distance of the stars from the cluster center and $r_{core}$ the core radius. We defined a rough cluster center, counting the cluster members in squares with sides of 0.3 parsec and taking the most populated square as the center. The coordinates of this center are $\alpha=18:18:41$ and $\delta=-13:47:23$. Then we calculated the projected cluster member density (number of members per parsec$^{-2}$) in concentric circles with increasing radius. To find the central density and core radius, we performed a non-linear fit between our distribution and the King's radial density profile. We found a central density ($\sigma_0$) of 140 $\pm$ 8 stars/parsec$^2$ and a projected core radius (r$_{core}$) equal to 1.40 $\pm$ 0.08 parsec (see Fig. \ref{rdp}). We repeated this procedure for the stars with a circumstellar disk and the X-ray sources separately. We again obtained a good agreement between the observed and the King's profile, with a core radius of 2.34 $\pm$ 0.44 parsec for the stars with a circumstellar disk and 1.30 $\pm$ 0.08 parsec for the X-ray sources. This result suggests that the distribution of stars with a circumstellar disk is less concentrated near the center of the cluster, where most of OB stars are, than the distribution of the X-ray sources. \par
	The radial density profile and the dynamical state of the cluster has been studied by \citet{Bona06}, but using only 2MASS data and a membership criterion based on the NIR colors. The good agreement between the observed and the King's density profile suggested to \citet{Bona06} that the internal region of NGC~6611 has reached some level of energy equipartition. Using the number of stars that they found in the core (120), the core radius (0.7 pc) and the definition of relaxation time: 
	
	\begin{equation}
	t_{relax}=\frac{N}{8lnN}\times \frac{r_{core}}{\sigma_{v}}   
	\label{trelax}
	\end{equation}
	
	where N in total number of cluster members and $\sigma_{v}$ the velocity dispersion, with the typical value of 3 Km s$^{-1}$,  \citep{Bin98}, \citet{Bona06} found the relaxation time to be 0.7 Myr. This value of $t_{relax}$ is about half the age of the cluster that they estimated (1.3 Myr), and this reinforced their hypothesis about the energy equipartition level reached in the cluster's inner region.\par
\begin{figure} []	
	\includegraphics[angle=0,width=8cm]{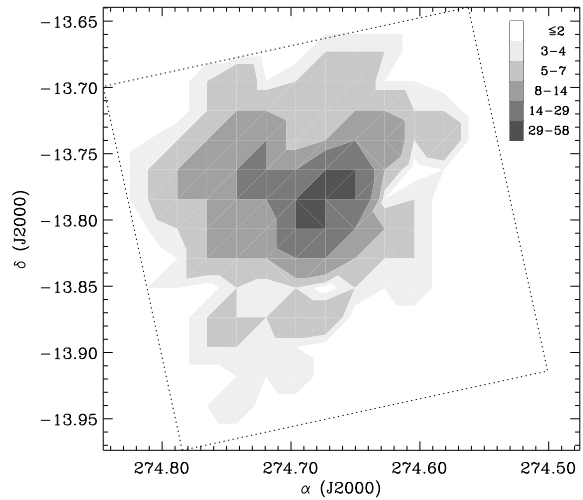}
	\caption{ Color map of cluster members in the ACIS FOV (delimited by the dotted line) in squares with sides of 0.3 pc.}
	\label{contorni}
	\end{figure}

	 Using the value of $r_{core}$ found in this work, and the value of $N$ for the core, estimated to be equal to 448 by our number of cluster members and the IMF found in the study about the Orion cluster of \citet{Mue02}, we found a $t_{relax}$ of $\sim 4.2$ Myr, greater than the upper limit of the age of the X-ray sources (3 Myr, see Sect. \ref{parameter}). The hypothesis that the core of the cluster has reached some level of energy equipartition is therefore not supported by our data. This is also confirmed by the significant asymmetry of the spatial distribution of cluster members, with very few cluster members toward the South-West (Fig. \ref{contorni}). \par

\section{Influence of the OB stars' radiation on the circumstellar disks.}
\label{photoevaporation}

	We investigated the effects of the UV radiation from the numerous OB stars of NGC~6611 on the evolution of the circumstellar disks of the lower-mass cluster members. To do this, we calculated the bolometric flux emitted by all the massive cluster members earlier than B5 and incident on the positions in the ACIS FOV of stars with excess in the K band. In order to take into account the inhomogeneous spatial distribution of the cluster, we compared the distribution of stars with disks with that of X-ray sources. We have not considered the absorption of the UV radiation of the hot stars due to the intracluster medium. Since the difference in extinction between the foreground stars and the X-ray sources is about A$_V=1.3$ and the UV radiation penetrates to a depth A$_V\sim 1$, this effect is probably important only for the cluster members far away from the center of NGC~6611, where most of the OB stars are. We also used the projected distances between the stars. We will discuss later the effects of this approximation on our results. \par

	\begin{figure*} []	
	\includegraphics[angle=0,width=8cm]{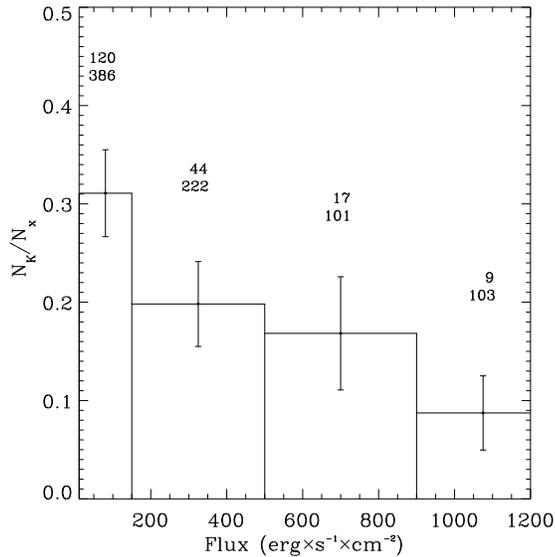}\par
	\caption{Histogram of the ratio between the number of the stars with excess in K band in the ACIS FOV (upper numbers) and the X-ray sources without circumstellar disk (lower numbers) vs. the incident flux from OB stars. We also show the error bars of the ratio.}
	\label{isto}
	\end{figure*}

	\begin{figure*} []	
	\includegraphics[angle=0,width=16cm]{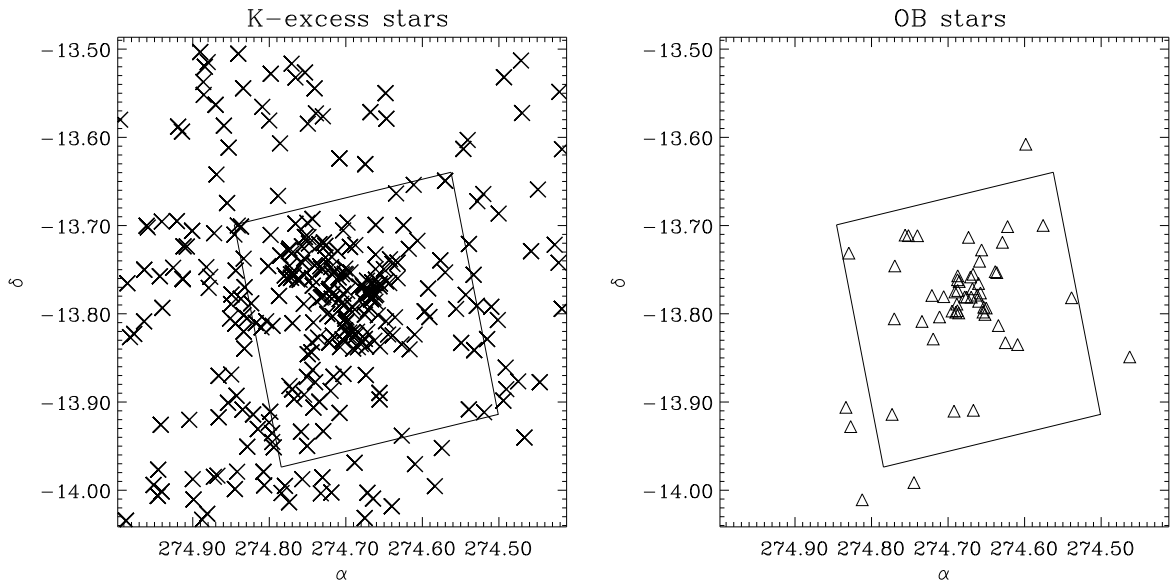}\par
	\caption{Spatial distribution of the stars with excess in the K band in the WFI FOV (left panel) and of the OB stars members of NGC~6611 (right panel). The rotated square marks the ACIS FOV.}
	\label{kobpos}
	\end{figure*}

	We have binned the values of incident fluxes in four ranges, and determined the ratio between the numbers of K-excess stars and  X-ray sources without a circumstellar disk for each range. This normalization is needed to account for the spatial structure of the cluster and it is based on the assumption that the structure is well traced by the X-ray sources (the richest group of members). The results are shown in Fig. \ref{isto}, where it is evident that the fraction of K-excess stars is significantly larger in the first bin (i.e. at low values of incident flux) than in the other bins. \par
	If we could take into account the absorption of UV radiation by the intracluster medium and if we could use the real distances between the stars, instead of the projected ones, we would obtain lower values of incident fluxes, so the first bin would become more populated, reinforcing our results. Our analysis is therefore conservative in this respect. The spatial distribution of the K-excess stars is then anti correlated with the spatial distribution of OB stars. \par
	The spatial distribution of stars in the WFI FOV with excess in the K band is shown in the left panel of Fig. \ref{kobpos}. It is evident that the cluster is more extended than the ACIS FOV, with star-formation events in the outer regions of the nebula, and more dispersed than the OB stars (right panel); the spatial distribution of the X-ray sources is shown in Fig. \ref{xsources}.
	
\subsection{Interpretation of the results}
\label{theend}

	We have found that the spatial distribution of the stars with a circumstellar disk in NGC~6611 is anti-correlated with the distribution of OB stars. We have shown that this is not due to energy equipartition. We have also verified that the possible alternative interpretation, that stars close to massive stars are older than the remaining population (therefore with more evolved disks) is not supported by their position in the optical color-magnitude diagrams, so there are no effects due to sequential star formation. \par
	We then explain the spatial distribution of stars with a circumstellar disk with the hypothesis that UV radiation from OB stars causes the photoevaporation of the disks close to massive stars, altering the evolution of the circumstellar disks. \par
	The difference between our result and that of \citet{Oli05}, who did not find any influence of massive stars on disc frequency for NGC~6611, is mainly due to the different members selected. Our X-ray based criterion allows us to also select WTTS, that are very difficult to identify with other methods. \par 
	The process of photoevaporation works from outside (where the gravitational attraction of the central star is smaller) inward, up to the gravitational radius ($r_g$), where the escape velocity is equal to the sound speed (see \citealt{Holle94}). For example, in a star with a mass of 0.2 $M_{\odot}$ with a circumstellar disk heated by the incident UV radiation to 1000 K (FUV regime), the gravitational radius is at 20 AU from the central stars; in the EUV regime, when the circumstellar disk can be heated up to 10$^4$ K by the incident UV radiation, the gravitational radius is at 2 AU. For distances from the central star smaller than the gravitational radius, the photoevaporated gas produces a stable atmosphere around the star. However, \citet{Woods96} have shown that if the evaporating gas reaches supersonic velocities in the region inside the gravitational radius, then the photoevaporation occurs even closer to the central stars (up to about 0.2$\times r_g$). If the outer regions of the circumstellar disk are photoevaporated, accretion cannot drain material in the inner region replenishing the material dissipated by the stellar radiation or  accreted on the star's surface. This process then may also cause the destruction of the inner region, responsible for the emission in NIR bands. \par
	We have used only the current position of the stars to calculate the flux from the massive stars incident on the circumstellar disk, and we have not considered the orbit of the stars during the evolution of the cluster. Because of the young age of NGC~6611, this can be relevant only for the stars in the core, while the stars in the outer region have not had enough time to pass near the center where more OB stars are clustered. However, the exact computation of the incident UV radiation, integrated in the orbit during the lifetime of the stars with a circumstellar disk is beyond the scope of this paper. Using N-body simulations of the evolution of clusters with a population between 100-1000 members, \citet{Ada06} claimed that the destruction of the circumstellar disks and young planetary systems are rare events. This result is not applicable in the case of NGC~6611, with a larger population (1122 members considering only masses $>0.4M_{\odot}$).  \par

\section{Summary}

	We have studied the young open cluster NGC~6611 to derive the spatial distribution of the stars with circumstellar disks and its correlation with the population of OB cluster members. We found that the stars with excess in K, due to the presence of the circumstellar disks, are relatively more frequent at larger distances from OB stars. We interpret this fact as due to a fast disk evaporation induced by the incident UV radiation emitted by the numerous massive stars of NGC~6611. \par
	We built a deep optical catalog compiled for the region of NGC~6611, using a WFI observation in bands B, V and I, and comprising 32308 optical sources with magnitudes V$\leq23^m$. Using our optical catalog, a CHANDRA/ACIS observation with 997 X-ray sources, and the 2MASS data in the WFI FOV, we built a multiband catalog comprising 38995 optical, NIR and X-ray sources. \par
	The catalog allowed us to improve our knowledge about NGC~6611. We estimated the distance of the cluster equal to about 1750 parsec, a value smaller than that of the work of H93 and B99, but similar to the one found in more recent works; we also confirmed the presence of an anomalous reddening law in the direction of NGC~6611, with $R_V=3.27$. Using the X-ray emission as a membership criterion, we found a large number of cluster members with masses $\leq 1$ $M_{\odot}$. We found stars with ages up to 3 Myr together with very young stars with ages of about 0.1 Myr. Most of the previous works identified only stars with ages less than 1 Myr. \par 
	We identified 342 sources with excesses in the NIR bands in the WFI FOV using suitable color indices, independent of the extinction. In the ACIS FOV, 47\% of the stars with circumstellar disks have an X-ray counterpart (94 X-ray sources with 192 stars with K band excess in the ACIS FOV).
	The stars with circumstellar disks and those that are X-ray sources form the most numerous and the deepest list of members of NGC~6611 (1122 cluster members in total). \par
	Using this member list we have studied the radial density profile of the cluster, finding a core radius of 1.40 $\pm$ 0.08 parsec.	The spatial distribution of the stars with a circumstellar disk is less concentrated near the cluster center, where most of the OB stars are, than the X-ray sources. The fraction of stars with disks increases with decreasing incident UV flux from early type stars. \par
	There are 194 stars with a circumstellar disk outside the ACIS FOV, showing that the Eagle Nebula is characterized by star-formation events. 
	
\begin{acknowledgements}
We thank the anonymous referee for useful comments. We acknowledge financial support from the Ministero dell'Universit\'a e Ricerca Scientifica e Tecnologica. This publication make use of the Two Micron All Sky Survey (2MASS), a joint project of the University of Massachusetts and the Infrared Processing and Analysis Center/California Institute of Technology, funded by the National Aeronautics and Space Administration and the National Science Foundation. 
\end{acknowledgements}
	
\appendix
\onecolumn
\section{Coefficients of the solution for the photometric calibration}

\label{aptable}

	\begin{table*}[!h]
	\centering
	\caption {Zero-points, coefficients of the colors and temporal terms for the CCDs, used in the transformation (\ref{mysolution}).}
	\vspace{0.2cm}

	\begin{tabular}{rrrrrrrrr}
	\hline
	\hline 
	CCD& Z & $\sigma_Z$ & A$_1$ & $\sigma_{A_1}$ & A$_2$ & $\sigma_{A_2}$ & $A_t$ & $\sigma_{A_t}$   \\
	\hline 
        $Filter$ $B$ &$$ &$$ &$$ &$$ &$$ &$$  &$$  &$$\\	
	$\#50$ &$1.008$ &$0.004$ &$-0.182$ &$0.015$ &$      $ &$    $  &$	$  &$      $\\
	$\#51$ &$0.561$ &$0.002$ &$-0.271$ &$0.006$ &$      $ &$    $  &$	$  &$      $\\
	$\#52$ &$0.637$ &$0.003$ &$-0.293$ &$0.011$ &$      $ &$    $  &$-0.0028$  &$0.0007$\\
	$\#53$ &$0.639$ &$0.002$ &$-0.334$ &$0.008$ &$0.03  $ &$0.02$  &$	$  &$      $\\
	$\#54$ &$0.583$ &$0.002$ &$-0.255$ &$0.008$ &$      $ &$    $  &$-0.0021$  &$0.0004$\\
	$\#55$ &$0.582$ &$0.002$ &$-0.333$ &$0.009$ &$      $ &$    $  &$-0.0042$  &$0.0004$\\
	$\#56$ &$0.531$ &$0.003$ &$-0.349$ &$0.013$ &$      $ &$    $  &$	$  &$      $\\
	$\#57$ &$0.405$ &$0.004$ &$-0.127$ &$0.023$ &$      $ &$    $  &$	$  &$      $\\
	\hline
	\hline 
        $Filter$ $V$ &$$ &$$ &$$ &$$ &$$ &$$  &$$  &$$\\	
	$\#50$ &$0.824$ &$0.002$ &$0.145^{**}$ &$0.007$ &$      $ &$   	$  &$      $  &$     $\\
	$\#51$ &$0.778$ &$0.002$ &$0.089^*   $ &$0.009$ &$      $ &$	$  &$      $  &$     $\\
	$\#52$ &$1.003$ &$0.001$ &$0.022^{**}$ &$0.005$ &$0.030 $ &$0.016$  &$-0.0028$  &$0.0004$\\
	$\#53$ &$0.956$ &$0.001$ &$0.084^*   $ &$0.004$ &$-0.003$ &$0.001$  &$      $  &$     $\\
	$\#54$ &$0.900$ &$0.002$ &$0.072^*   $ &$0.005$ &$      $ &$	$  &$-0.0035$  &$0.0004$\\
	$\#55$ &$0.900$ &$0.001$ &$0.056^*   $ &$0.005$ &$      $ &$	$  &$-0.0029$  &$0.0004$\\
	$\#56$ &$0.737$ &$0.004$ &$0.155^{**}$ &$0.014$ &$-0.04 $ &$0.02 $  &$      $  &$     $\\
	$\#57$ &$0.740$ &$0.004$ &$0.082^{**}$ &$0.021$ &$      $ &$	$  &$      $  &$     $\\
	\multicolumn{7}{l}{We used the color $B-V$ ($**$) or the color $V-I$ ($*$).} &$$  &$$\\
	\hline
	\hline 
        $Filter$ $I$ &$$ &$$ &$$ &$$ &$$ &$$  &$$  &$$\\	
	$\#50$ &$1.728$ &$0.004$ &$-0.112$ &$0.006$ &$      $ &$    $  &$	  $  &$     $\\
	$\#51$ &$1.836$ &$0.003$ &$-0.142$ &$0.006$ &$      $ &$    $  &$0.0040 $  &$0.0020$\\
	$\#52$ &$1.979$ &$0.002$ &$-0.095$ &$0.004$ &$-0.016$ &$0.006$  &$-0.0024$  &$0.0005$\\
	$\#53$ &$1.921$ &$0.002$ &$-0.122$ &$0.006$ &$      $ &$    $  &$-0.0070$  &$0.0020$\\
	$\#54$ &$1.872$ &$0.003$ &$-0.136$ &$0.005$ &$      $ &$    $  &$-0.0010$  &$0.0006$\\
	$\#55$ &$1.890$ &$0.002$ &$-0.140$ &$0.004$ &$      $ &$    $  &$-0.0016$  &$0.0004$\\
	$\#56$ &$1.763$ &$0.004$ &$-0.152$ &$0.005$ &$      $ &$    $  &$	  $  &$     $\\
	$\#57$ &$2.442$ &$0.004$ &$-0.182$ &$0.010$ &$      $ &$    $  &$	  $  &$     $\\
	\hline
	\hline
	\multicolumn{9}{l}{A$_n$ is the coefficient of the color term with exponent equal to n.}
	\end{tabular}
	\label{coef1}
       \end{table*}

       \begin{table*}[]
	\centering
	\caption {Coefficients of the position's polynomials used for the photometric calibration in all the CCDs. }
	\vspace{0.2cm}
	\tabcolsep 0.15truecm
	\begin{tabular}{rrrrrrrrrrrrrrr}
	\hline
	\hline
	$$  \vline & \multicolumn{4}{c}{Linear terms} \vline & \multicolumn{6}{c}{Quadratic terms} \vline & \multicolumn{4}{c}{Cubic terms}   \\
	\hline
	\hline 
	CCD& $A_{X}$&$\sigma_{A_X}$ &$A_{Y}$&$\sigma_{A_Y}$ &$A_{X^2}$&$\sigma_{A_{X^2}}$ &$A_{Y^2}$&$\sigma_{A_{Y^2}}$ &$A_{XY}$&$\sigma_{A_{XY}}$ &$A_{X^3}$&$\sigma_{A_{X^3}}$ &$A_{Y^3}$&$\sigma_{A_{Y^3}}$   \\

	\hline 
        $Filter$ $B$ &$$ &$$ &$$ &$$ &$$ &$$&$$ &$$ &$$ &$$ &$$ &$$ &$$  &$$\\
	$\#50$ &$-1.058$ &$0.520$ &$-0.067$ &$0.017$ &$0.64 $ &$0.41 $ &$      $ &$     $ &$0.057$ &$0.014$ &$      $ &$     $  &$       $  &$      $\\
	$\#51$ &$0.047 $ &$0.011$ &$-0.023$ &$0.003$ &$     $ &$     $ &$      $ &$     $ &$     $ &$     $ &$-0.007$ &$0.002$  &$       $  &$      $\\
	$\#52$ &$      $ &$     $ &$      $ &$     $ &$     $ &$     $ &$      $ &$     $ &$0.016$ &$0.003$ &$-0.011$ &$0.002$  &$-0.0016$  &$0.0002$\\
	$\#53$ &$-0.068$ &$0.008$ &$      $ &$     $ &$     $ &$     $ &$      $ &$     $ &$     $ &$     $ &$0.016 $ &$0.002$  &$       $  &$      $\\
	$\#54$ &$-0.037$ &$0.034$ &$      $ &$     $ &$     $ &$     $ &$      $ &$     $ &$     $ &$     $ &$      $ &$     $  &$       $  &$      $\\
	$\#55$ &$0.081 $ &$0.029$ &$0.017 $ &$0.006$ &$-0.04$ &$0.01 $ &$      $ &$     $ &$     $ &$     $ &$      $ &$     $  &$       $  &$      $\\
	$\#56$ &$0.151 $ &$0.085$ &$0.027 $ &$0.015$ &$-0.12$ &$0.09 $ &$-0.004$ &$0.003$ &$0.012$ &$0.006$ &$      $ &$     $  &$       $  &$      $\\
	$\#57$ &$      $ &$     $ &$      $ &$     $ &$-0.22$ &$0.11 $ &$      $ &$     $ &$     $ &$     $ &$0.104 $ &$0.037$  &$       $  &$      $\\

	\hline
	\hline
        $Filter$ $V$ &$$ &$$ &$$ &$$ &$$ &$$&$$ &$$ &$$ &$$ &$$ &$$ &$$  &$$\\
	$\#50$ &$0.085 $ &$0.055$ &$-0.098$ &$0.025$ &$-0.033$ &$0.020$ &$0.0275 $ &$0.0070$ &$0.013$ &$0.008$ &$      $ &$     $  &$          $  &$      $\\
	$\#51$ &$0.275 $ &$0.029$ &$-0.048$ &$0.018$ &$-0.087$ &$0.011$ &$0.0171 $ &$0.0054$ &$0.021$ &$0.005$ &$-0.007$ &$0.002$  &$       $  &$      $\\
	$\#52$ &$0.021 $ &$0.008$ &$-0.006$ &$0.003$ &$      $ &$     $ &$       $ &$      $ &$0.016$ &$0.003$ &$-0.005$ &$0.002$  &$-0.0014$  &$0.0002$\\
	$\#53$ &$      $ &$     $ &$      $ &$     $ &$-0.156$ &$0.008$ &$-0.0041$ &$0.0006$ &$0.003$ &$0.002$ &$0.089 $ &$0.003$  &$       $  &$      $\\
	$\#54$ &$-0.067$ &$0.010$ &$      $ &$     $ &$      $ &$     $ &$0.0216 $ &$0.0031$ &$     $ &$     $ &$0.020 $ &$0.003$  &$-0.0046$  &$0.0008$\\
	$\#55$ &$      $ &$     $ &$0.047 $ &$0.004$ &$      $ &$     $ &$       $ &$      $ &$     $ &$     $ &$-0.003$ &$0.002$  &$-0.0013$  &$0.0002$\\
	$\#56$ &$0.114 $ &$0.021$ &$0.042 $ &$0.010$ &$      $ &$     $ &$       $ &$      $ &$     $ &$     $ &$-0.021$ &$0.005$  &$-0.0010$  &$0.0006$\\
	$\#57$ &$0.192 $ &$0.053$ &$      $ &$     $ &$-0.053$ &$0.020$ &$-0.0307$ &$0.0223$ &$     $ &$     $ &$      $ &$     $  &$       $  &$      $\\

	\hline
	\hline
        $Filter$ $I$ &$$ &$$ &$$ &$$ &$$ &$$&$$ &$$ &$$ &$$ &$$ &$$ &$$  &$$\\
	$\#50$ &$0.251 $ &$0.065$ &$      $ &$     $ &$-0.07$ &$0.02 $ &$      $ &$     $ &$      $ &$     $ &$      $ &$     $  &$       $  &$      $\\
	$\#51$ &$0.285 $ &$0.058$ &$-0.09 $ &$0.03 $ &$-0.10$ &$0.02 $ &$0.016 $ &$0.008$ &$      $ &$     $ &$      $ &$     $  &$       $  &$      $\\ 
	$\#52$ &$      $ &$     $ &$      $ &$     $ &$0.02 $ &$0.01 $ &$-0.015$ &$0.003$ &$0.010 $ &$0.004$ &$-0.010$ &$0.006$  &$0.0012 $  &$0.0008$\\
	$\#53$ &$-0.046$ &$0.004$ &$      $ &$     $ &$     $ &$     $ &$      $ &$     $ &$      $ &$     $ &$      $ &$     $  &$-0.0039 $  &$0.0020$\\
	$\#54$ &$      $ &$     $ &$0.03  $ &$0.01 $ &$-0.03$ &$0.01 $ &$-0.003$ &$0.002$ &$      $ &$     $ &$      $ &$     $  &$       $  &$      $\\
	$\#55$ &$      $ &$     $ &$      $ &$     $ &$0.02 $ &$0.01 $ &$0.029 $ &$0.004$ &$-0.014$ &$0.003$ &$-0.008$ &$0.007$  &$-0.0049 $  &$0.0009$\\
	$\#56$ &$0.216 $ &$0.078$ &$0.08  $ &$0.01 $ &$-0.14$ &$0.08 $ &$-0.011$ &$0.003$ &$0.016 $ &$0.006$ &$      $ &$     $  &$       $  &$      $\\
	$\#57$ &$-1.185$ &$0.370$ &$-0.14 $ &$0.03 $ &$1.03 $ &$0.18 $ &$      $ &$     $ &$      $ &$     $ &$-0.260$ &$0.108$  &$0.0067 $  &$0.0017$\\
	\hline
	\hline
	\end{tabular}
	\label{coef2}
 	\end{table*}


\twocolumn
\addcontentsline{toc}{section}{\bf Bibliografia}
\bibliographystyle{apj}
\bibliography{biblio}

\end{document}